**Does the specification of uncertainty hurt the progress of scientometrics?**

*Sir,*

In a recent paper, entitled "*Caveats* for using statistical significance tests in research assessments," Schneider (2013) focuses on "Leydesdorff and co-authors"—notably Opthof & Leydesdorff (2010)—as papers that misuse of statistics in the social sciences. Schneider considers our practice as "not necessarily inferior to those of most others." His main objection is that we did not discuss the endlessly repeated debate about significance testing among statisticians (e.g., Cohen, 1994; McClosky, 1985). However, we focused on scientometric problems with indicators that, in our opinion, were mistaken because of (*i*) an error against the order of operations (Lundberg, 2007), and (*ii*) insufficient specificity of the reference sets used for the normalization (Leydesdorff & Opthof, 2010; cf. Bornmann *et al.*, 2008). The semi-industrial production of these indicators generated noise in evaluations (e.g., CWTS, 2008, 2010; Spaan, 2010).

The "crown indicator" of CWTS (*CPP/FCSm*) was based on a quotient between two gross averages instead of averaging the ratios of observed versus expected numbers of citations (Gingras & Larivière, 2011). Unlike a mean, a single quotient does not allow for any specification of uncertainty. CWTS, which has abandoned this indicator in response to our critique, advocated at the time the use of a rule of thumb: a deviation of 0.2 from the world average of 1.0 was considered significant (Van Raan, 2005, at p. 7; cf. CWTS, 2008, p. 7; Glänzel, 1992; Schubert & Glänzel, 1983). Glänzel (2010) further explained that this "rule of thumb" is based on the Central Limit Theorem. However, this theorem is invalid for finite sets with skewed distributions in scientometric evaluation studies.[1] In my opinion, one should use statistics properly for the specification of uncertainty, and follow the order of operations.

The error bars in Opthof & Leydesdorff (2010) were at the time still based on using means of skewed distributions (Seglen, 1992). However, we used Dunn's test as non-parametric statistics. When Bornmann & Mutz (2011) thereafter entered this debate to plead for using percentile ranks, we joined forces with this team and developed a full set of indicators based on percentile ranks (Leydesdorff, Bornmann, Opthof, & Mutz, 2011). The following criteria for scientometric indicators were specified (at pp. 1371f.):

1. A citation-based indicator must be defined so that the choice of the reference set(s) (e.g, journals, fields) can be varied by the analyst independently of the question of the evaluation scheme. In other words, these two dimensions of the problem (the normative and the analytical ones) have to be kept separate;
2. The citation indicator should accommodate various evaluation schemes, for example, by funding agencies. Some agencies may be interested in the top-1% (e.g., National Science Board, 2010) while others may be interested in whether papers based on research funded by a given agency perform significantly better than comparable non-funded ones (e.g., Bornmann *et al.*, 2010);
3. The indicator should allow productivity to be taken into account. One should, for example, be able to compare two papers in the 39[th] percentile with a single paper in the 78[th] percentile (with or without weighting the differences in rank in an evaluation scheme as specified under 2.);
4. The indicator should provide the user, among other things, with a relatively straightforward criterion for the ranking (for example, a percentage of a maximum) that can then be tested for its statistical significance in relation to comparable (sets of) papers;
5. It should be possible to calculate the statistical error of the measurement.

---
[1] Schubert & Glänzel (1983) based their reasoning on normal distributions (Glänzel, 2010).

In my opinion, these are theoretical conclusions since they are not dependent on the use of one statistics or another.

Schneider's (2013) core message can be found at the end of this article where he states that "significance tests, confidence intervals, and standard errors" cannot make a decision for us. Unlike some university departments and private companies that produce indicator statistics for legitimating decisions, I am not in the business of providing management information, but we criticized the publication of indicators lacking specification of error. Indicators with sometimes three decimals, but without specification of uncertainty and graphs without error bars, in our opinion, should be mistrusted because the reader may easily draw the wrong conclusions while ranks are tied. Even if error is specified, the validity of using citation measures in evaluations as indicators of quality remains an issue of theoretical concern (e.g., Leydesdorff & Amsterdamska, 1991).

I agree with Schneider (2013) insofar as he proposes to develop further statistical instruments (such as effect sizes). Bornmann & Leydesdorff (in press) responded to Schneider's (2012) previous critique by adding effect sizes to the online tests of the Leiden Rankings (at http://www.leydesdorff.net/leiden11/leiden11.xls) and the Scimago Excellence indicator (at http://www.leydesdorff.net/scimago11/scimago11.xls; cf. Cohen, 1988). Against Schneider (2013), it seems to me that decision makers are served better with a specification of uncertainty—given the current state of the art—than without statistical information. In several studies, for example, the failure to distinguish between "excellent" and "good" research using scientometric indicators was shown by using statistics (Bornmann, Leydesdorff, Van den Besselaar, 2009; Opthof & Leydesdorff, 2011). Such counter-intuitive results raise further research questions and can thus be fruitful heuristically. Policy-relevant questions can also be raised. Schneider (2013), however, argues on meta-theoretical grounds against the specification of uncertainty because, in his opinion, the presence of statistics (including confidence levels and effect sizes) would legitimate decision-making. I disagree: uncertainty can also be used for opening a debate. In my opinion, scientometric results in which error bars are suppressed for meta-theoretical reasons should not be trusted.

Loet Leydesdorff [i]

**References**
Bornmann, L., & Leydesdorff, L. (in press). Statistical Tests and Research Assessments: A comment on Schneider (2012). *Journal of the American Society for Information Science and Technology*.

---

[i] Amsterdam School of Communications Research (ASCoR), University of Amsterdam, Kloveniersburgwal 48, 1012 CX Amsterdam, The Netherlands; loet@leydesdorff.net; http://www.leydesdorff.net.